\documentclass[conference]{IEEEtran}
\IEEEoverridecommandlockouts

\usepackage{cite}
\usepackage{amsmath,amssymb,amsfonts}
\usepackage{graphicx}
\usepackage{subfigure}
\usepackage{threeparttable}
\usepackage{setspace}
\usepackage{makecell}
\usepackage[table,xcdraw]{xcolor}
\usepackage{tabularx}
\usepackage{textcomp}
\usepackage{xcolor}
\usepackage{stfloats}
\usepackage{amsfonts}
\usepackage{amssymb}
\usepackage{amsthm}
\usepackage{cite}
\usepackage{float}
\usepackage{color}
\usepackage{stfloats,fancyhdr}
\usepackage{amsmath,bm}
\usepackage{algorithm}
\usepackage{changepage}
\usepackage[normalem]{ulem}
\usepackage{balance}
\usepackage{graphicx}
\usepackage{bbm}
\usepackage{graphicx}
\usepackage{bbm}

\usepackage{threeparttable}
\usepackage{algpseudocode}
\usepackage{setspace}

\newtheorem{problem}{Problem}

\newtheorem{definition}{Definition}

\usepackage{mathtools}
\mathtoolsset{showonlyrefs=true}    
\usepackage{comment}

\usepackage{geometry} 

\title{\huge Structure-Enhanced Deep Reinforcement Learning for Optimal Transmission Scheduling \vspace{-0.5cm} \\
\thanks{*Wanchun Liu is the corresponding author.}
}

\author{
	\IEEEauthorblockN{Jiazheng Chen$^1$, Wanchun Liu*$^1$, Daniel E.\ Quevedo$^2$, Yonghui Li$^1$, and Branka Vucetic$^1$}
\IEEEauthorblockA{$^1$School of Electrical and Information Engineering, The University of Sydney, Australia \\
$^2$School of Electrical Engineering and Robotics, Queensland University of Technology, Brisbane, Australia.\\	
$^1$Emails: \{jiazheng.chen, wanchun.liu, yonghui.li, branka.vucetic\}@sydney.edu.au. $^2$Email: dquevedo@ieee.org.} 
\vspace{-0.9cm}
}
\begin{document}
\maketitle

\begin{abstract}
Remote state estimation of large-scale distributed dynamic processes plays an important role in Industry 4.0 applications. 
In this paper, by leveraging the theoretical results of structural properties of optimal scheduling policies,
we develop a structure-enhanced deep reinforcement learning (DRL) framework for optimal scheduling of a multi-sensor remote estimation system to achieve the minimum overall estimation mean-square error (MSE).
In particular, we propose a structure-enhanced action selection method, which tends to select actions that obey the policy structure. This explores the action space more effectively and enhances the learning efficiency of DRL agents.
Furthermore, we introduce a structure-enhanced loss function to add penalty to actions that do not follow the policy structure. The new loss function guides the DRL to converge to the optimal policy structure quickly.
Our numerical results show that the proposed structure-enhanced DRL algorithms can save the training time by 50\% and reduce the remote estimation MSE by 10\% to 25\%, when compared to benchmark DRL algorithms.
\end{abstract}


\section{Introduction} \label{sec:intro}
Wireless networked control systems (WNCSs) consisting of distributed sensors, actuators, controllers, and plants are a key component for Industry 4.0 and have been widely applied in many areas such as industrial automation, vehicle monitoring systems, building automation and smart grids \cite{Park2018WNCS}. In particular, providing high-quality real-time remote estimation of dynamic system states plays an important role in ensuring control performance and stability of WNCSs \cite{Huang2020UpDown}.
For large-scale WNCSs, transmission scheduling of wireless sensors over limited bandwidth needs to be properly designed to guarantee remote estimation performance.

There are many existing works on the transmission scheduling of WNCSs. 
In \cite{gatsis2015opportunistic}, the optimal scheduling problem of a multi-loop WNCS with limited communication resources was investigated for minimizing the transmission power consumption under a WNCS stability constraint.
In~\cite{han2017optimal,wu2018optimalmulti}, optimal sensor scheduling problems of remote estimation systems were investigated for achieving the best overall estimation mean-square error (MSE). In particular, dynamic decision-making problems were formulated as Markov decision processes (MDPs), which can be solved by classical  methods, such as policy and value iterations.
However, the conventional model-based solutions are not feasible in large-scale scheduling problems because of the curse of dimensionality caused by high-dimensional state and action spaces.\footnote{In~\cite{gatsis2015opportunistic,han2017optimal,wu2018optimalmulti}, only the optimal scheduling of two-sensor systems has been solved effectively by the proposed methods.} 

In recent years, deep reinforcement learning (DRL) has been developed to deal with large MDPs by using deep neural networks as function approximators \cite{sutton2018reinforcement, zhao2022deep}. 
Some works \cite{leong2020DRL, liu2021drlscheduling} have used the deep Q-network (DQN), a simplest DRL method, to solve multi-sensor-multi-channel scheduling problems in different remote estimation scenarios. In particular, the sensor scheduling problems for systems with 6 sensors have been solved effectively, providing significant performance gains over heuristic methods in terms of estimation quality. More recent work~\cite{pang2022drl} has introduced DRL algorithms with an actor-critic structure to solve the scheduling problem at a much larger scale (that cannot be handled by the DQN). However, existing works merely use the general DRL frameworks to solve specific scheduling problems, without questioning what features make sensor scheduling problems different from other MDPs. Also, we note that a drawback of general DRL is that it often cannot perform policy exploration effectively for specific tasks\cite{guo2019exploration}, which can lead to local minima or even total failure. Thus, the existing DRL-based solutions could be far from optimal.

Some recent works have shown that the optimal scheduling policies of remote estimation systems commonly have a ``threshold structure" \cite{wu2020optimalmulti, wu2018optimalmulti}, which means that there exist switching boundaries dividing the state space into multiple regions for different scheduling actions. In other words, the optimal policy has a structure where the action only changes at the switching boundaries of the state space.
In~\cite{hsu2017threshold}, the threshold structure of an optimal scheduling policy has also been identified in a sensor scheduling system for minimizing the average age of information (AoI).
Although these theoretical works have derived the structures of optimal policies, there is no existing work in the open literature utilizing the structural properties to guide DRL algorithms. 
This knowledge gap motivates us to investigate how knowledge of the optimal policy structure can be used to redesign DRL algorithms that find the optimal policy more effectively.


\underline{\emph{Contributions.}} In this paper, we develop novel structure-enhanced DRL algorithms for solving the optimal scheduling problem of a remote estimation system. Given the most commonly adopted DRL frameworks for scheduling, i.e., DQN and deep deterministic policy gradient (DDPG), we propose structure-enhanced DQN and DDPG algorithms.
In particular, we design a structure-enhanced action selection method, which tends to select actions that obey the threshold structure. Such an action selection method can explore the action space more effectively and enhance the learning efficiency of DRL agents.
Furthermore, we introduce a structure-enhanced loss function to add penalty to actions that do not follow the threshold structure. The new loss function guides the DRL to converge to the optimal policy structure quickly.
Our numerical results show that the proposed structure-enhanced DRL algorithms can save the training time by 50\%, while reducing the remote estimation MSE by 10\% to 25\% when compared to benchmark DRL algorithms.

\vspace{-0.2cm}
\section{System Model} \label{sec: sys}
\begin{figure}[t]
    \centering
    \includegraphics[width=0.90\linewidth]{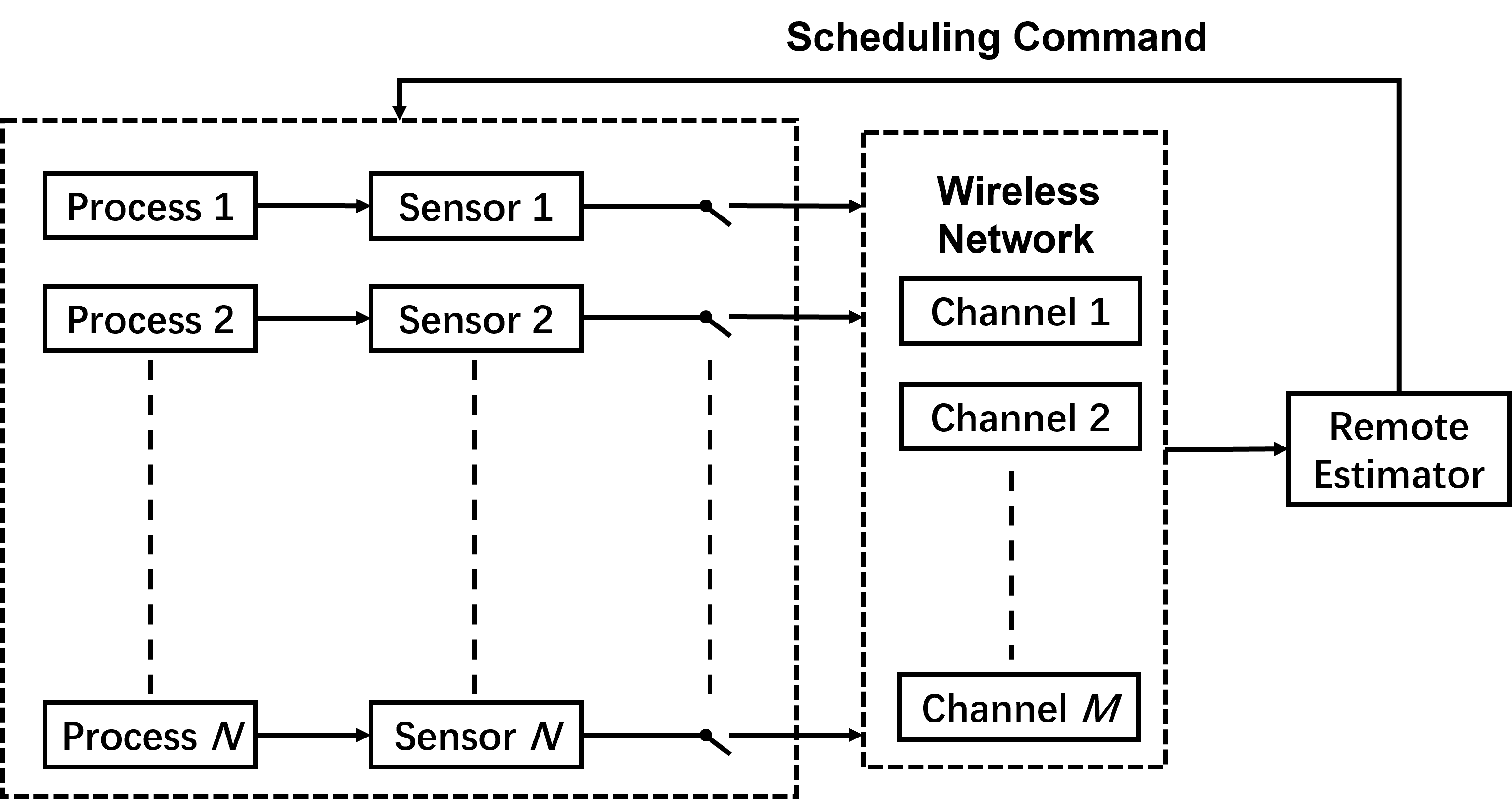}
    \vspace{-0.2cm}
    \caption{Remote state estimation system with $N$ processes and $M$ channels.}
    \label{fig:system_model}
    \vspace{-0.4cm}
\end{figure}

We consider a remote estimation system with $N$ dynamic processes each measured by a sensor, which pre-processes the raw measurements and sends its state estimates to a remote estimator through one of $M$ wireless channels, as illustrated in Fig.~\ref{fig:system_model}.

\subsection{Process Model and Local State Estimation}
Each dynamic process $n$ is modeled as a discrete linear time-invariant (LTI) system as \cite{leong2020DRL,liu2021remoteMF, liu2022stability}
\begin{equation}\label{eq:LTI}
\begin{aligned}
    \mathbf{x}_{n,k+1} & = \mathbf{A}_{n} \mathbf{x}_{n, k} + \mathbf{w}_{n, k},  \\
    \mathbf{y}_{n, k} & = \mathbf{C}_{n} \mathbf{x}_{n, k} + \mathbf{v}_{n, k},
\end{aligned}
\end{equation}
where $\mathbf{x}_{n, k} \in \mathbb{R}^{l_{n}}$ is  process $n$'s state at time $k$, and  $\mathbf{y}_{n, k} \in \mathbb{R}^{e_{n}}$ is the state measurement of the sensor $n$, 
$\mathbf{A}_{n} \in \mathbb{R}^{l_{n} \times l_{n}}$ and $\mathbf{C}_{n} \in \mathbb{R}^{e_{n} \times l_{n}}$ are the system matrix
and the measurement matrix, respectively, $\mathbf{w}_{n, k} \in \mathbb{R}^{l_{n}}$ and $\mathbf{v}_{n, k} \in \mathbb{R}^{e_{n}}$ are the process disturbance and the measurement noise modeled as independent and identically distributed (i.i.d) zero-mean Gaussian random vectors $\mathcal{N}(\mathbf{0},\mathbf{W}_{n})$ and $\mathcal{N}(\mathbf{0},\mathbf{V}_{n})$, respectively. \emph{We assume that the spectral radius of $\mathbf{A}_n, \forall n$, is greater than one, which means that the dynamic processes are unstable, making the remote estimation problem more interesting.}

Due to the imperfect state measurement in \eqref{eq:LTI}, sensor $n$ executes a classic Kalman filter to pre-process the raw measurement and generate state estimate $\mathbf{x}^s_{n,k}$ at each time $k$~\cite{liu2021remoteMF}.
Sensor $n$ sends $\mathbf{x}^s_{n,k}$ to the remote estimator (not $\mathbf{y}_{n,k}$) as a packet, once scheduled.
The local state estimation error covariance matrix is defined as 
\begin{equation}\label{eq:local_P}
    \mathbf{P}_{n, k}^{s} \triangleq \mathbb{E} \left[ (\mathbf{x}_{n, k}^{s} - \mathbf{x}_{n, k}) (\mathbf{x}_{n, k}^{s} - \mathbf{x}_{n, k})^{\top} \right].
\end{equation}
We note that local Kalman filters are commonly assumed to operate in the steady state mode in the literature (see \cite{liu2021remoteMF} and references therein\footnote{The $n$th local Kalman filter is stable if and only if that $(\mathbf{A}_{n},\mathbf{C}_{n})$ is observable and $(\mathbf{A}_{n}, \sqrt{\mathbf{W}_{n}})$ is controllable.}), which means the error covariance matrix converges to a constant, i.e.,  $\mathbf{P}_{n, k}^{s} = \bar{\mathbf{P}}_{n}, \forall k,n$.

\subsection{Wireless Communications and Remote State Estimation}
There are $M$ wireless channels (e.g., subcarriers) for the $N$ sensors' transmissions, where $M \leq N$.
We consider independent and identically distributed (i.i.d.) block fading channels, where the channel quality is fixed during each packet transmission and varies packet by packet, independently.
Let the $N\times M$ matrix $\mathbf{H}_{k}$ denote the channel state of the system at time $k$, where the element in the $n$th row and $m$th column  $h_{n,m,k} \in \mathcal{H} \triangleq \left\{ 1, 2, \dots, \bar{h} \right\}$ represents the channel state between sensor $n$ and the remote estimator at channel $m$. 
In particular, there are $\bar{h}$ quantized channel states in total. 
The packet drop probability at channel state $i \in \mathcal{H}$ is $p_i$. Without loss of generality, we assume that $p_{1} \geq p_{2} \geq \dots \geq p_{\bar{h}}$.
The instantaneous channel state $\mathbf{H}_{k}$ is available at the remote estimator based on standard channel estimation methods.
The distribution of $h_{n,m,k}$ is given~as
\begin{equation}\label{eq:q}
\operatorname{Pr}(h_{n,m,k}=i) = q^{(n,m)}_{i}, \forall k,
\end{equation}
where $\sum_{i=1}^{\bar{h}} q^{(n,m)}_{i} =1, \forall n,m$.

Due to the limited communication channels, only $M$ out of $N$ sensors can be scheduled at each time step. 
Let $a_{n,k} \in \{0, 1, 2, \dots, M\}$ represent the channel allocation for sensor $n$ at time $k$, where
\begin{equation}\label{eq:action}
    a_{n,k} = \left\{
    \begin{array}{ll}
        0  & \text{if sensor $n$ is not scheduled} \\
        m  & \text{if sensor $n$ is scheduled to channel $m$.}
    \end{array}
    \right.
\end{equation}
In particular, we assume that each sensor can be scheduled to at most one channel and that each channel is assigned to one sensor \cite{pang2022drl}. Then, the constraints on $a_{n,k}$ are given as
    \begin{equation}\label{eq: action constraint}
        \sum_{m=1}^{M} \boldsymbol{\mathbbm{1}}\left( a_{n,k} = m \right) \leq 1, \quad 
        \sum_{n=1}^{N} \boldsymbol{\mathbbm{1}}\left( a_{n,k} = m \right) = 1, 
    \end{equation} where $\boldsymbol{\mathbbm{1}} (\cdot)$ is the indicator function.

Considering schedule actions and packet dropouts, sensor $n$'s estimate may not be received by the remote estimator in every time slot. We define the packet reception indicator as
\begin{equation*}
\eta_{n, k} = \left\{
\begin{array}{ll}
1, & \text{if sensor $n$'s packet is received at time $k$}\\
0, & \text{otherwise}.
\end{array}
\right.
\end{equation*} 
Considering the randomness of $\eta_{n,k}$ and assuming that the remote estimator performs state estimation at the beginning of each time slot, the optimal estimator in terms of estimation MSE is given as
\begin{align}\label{eq:x_hat}
    \hat{\mathbf{x}}_{n, k+1} & = \left\{
                        \begin{array}{ll}
                             \mathbf{A}_n\mathbf{x}_{n, k}^{s}, & {\text{if $\eta_{n,k} = 1$}}\\
                             \mathbf{A}_n \hat{\mathbf{x}}_{n, k}, & {\text{otherwise}}
                        \end{array}
                    \right.
\end{align}
where $\mathbf{A}_n$ is the system matrix of process $n$ defined under \eqref{eq:LTI}.
If sensor $n$'s packet is not received, the remote estimator propagates its estimate in the previous time slot to estimate the current state.
From \eqref{eq:LTI} and \eqref{eq:x_hat}, we derive the estimation error covariance as
\begin{align}
  \mathbf{P}_{n, k+1} & \triangleq \mathbb{E} \left[ \left(\hat{\mathbf{x}}_{n, k} - \mathbf{x}_{n, k}\right) \left(\hat{\mathbf{x}}_{n, k} - \mathbf{x}_{n, k} \right)^{\top} \right] \\
& = \left \{
\begin{array}{ll}
\mathbf{A}_n\bar{\mathbf{P}}_{n}\mathbf{A}^\top_n + \mathbf{W}_n, & {\text{if $\eta_{n,k} = 1$}}\\
\mathbf{A}_n{\mathbf{P}}_{n,k}\mathbf{A}^\top_n + \mathbf{W}_n, & {\text{otherwise}},
\end{array} \label{eq:remote MSE}
\right. 
\end{align}
where $\bar{\mathbf{P}}_{n}$ is the local estimation error covariance of sensor $n$ defined under \eqref{eq:local_P}.

Let $\tau_{n,k}\in \{1,2,\dots\}$ denote the age-of-information (AoI) of sensor $n$ at time $k$, which measures the
amount of  time elapsed since the latest sensor packet was received. Then, we have
\begin{equation}\label{eq:tau}
\tau_{n,k+1} =\begin{cases}
1 & {\text{if $\eta_{n,k} = 1$}}\\
\tau_{n,k}+1 & {\text{otherwise.}}
\end{cases}
\end{equation}
If a sensor is frequently scheduled at good channels, the corresponding average AoI is small. However, due to the scheduling constraint~\eqref{eq: action constraint}, this is often not possible.

From \eqref{eq:remote MSE} and \eqref{eq:tau}, the error covariance can be simplified as
\begin{equation}\label{eq:MSE}
    \mathbf{P}_{n,k} = f_{n}^{\tau _{n,k}}(\bar{\mathbf{P}}_{n}),
\end{equation}
where
$f_{n}(\mathbf{X}) = \mathbf{A}_{n} \mathbf{X} \mathbf{A}_{n}^{\top} + \mathbf{W}_{n}$ and $f^{\tau+1}_{n}(\cdot) =f_n(f^{\tau}_{n}(\cdot))$. It has been proved that the estimation MSE, i.e., $\operatorname{Tr}(\mathbf{P}_{n,k})$, \emph{monotonically increases with the AoI state $\tau_{n,k}$~\cite{liu2021remoteMF}}.

\section{Problem Formulation and Threshold structure} \label{sec: problem}
In this paper, we aim to find a dynamic scheduling policy $\pi(\cdot)$ that uses the AoI states of all sensors, as well as the channel states to minimize the expected  total discounted estimation MSE of all $N$ processes over the infinite time horizon.
\begin{problem}\label{pro1}
\begin{equation}
    \max_{\pi} \lim_{T \to \infty} \mathbb{E}^\pi \left[ \sum_{k=1}^{T} \sum_{n=1}^N -\gamma^k \operatorname{Tr}(\mathbf{P}_{n,k}) \right],
\end{equation}
where $\mathbb{E}^\pi$ is the expected average value when adopting the policy $\pi$ and $\gamma<1$ is a discount factor.
\end{problem}

Problem~\ref{pro1} is a sequential decision-making problem with the Markovian property. This is because the expected estimation MSE only depends on the AoI state $\tau_{n,k}$ in \eqref{eq:MSE}, which has the Markovian property~\eqref{eq:tau}, and the channel states are i.i.d..
Therefore, we formulate Problem~\ref{pro1} as a Markov decision problem (MDP). 

\subsection{MDP Formulation}\label{sec:MDP}
We define the four elements of the MDP as below.

1) The state of the MDP is defined as $\mathbf{s}_k \triangleq \left( \bm{\tau}_k, \mathbf{H}_k \right) \in \mathbb{N}^N \times \mathcal{H}^{M \times N}$, where $\bm{\tau}_k = (\tau_{1,k}, \tau_{2,k}, \dots, \tau_{N,k}) \in \mathbb{N}^N$ is the AoI state vector. Thus, $\mathbf{s}_k$ takes into account both the AoI and channel states.
    
2) The overall schedule action of the $N$ sensors is defined as $\mathbf{a}_k = (a_{1,k}, a_{2,k}, \dots, a_{N,k}) \in \left\{ 0, 1, 2, \dots, M \right\}^N$ under the constraint~\eqref{eq: action constraint}. There are $N!/(N-M)!$ actions in total.
The stationary policy $\pi(\cdot)$ is a mapping between the state and the action, i.e., $\mathbf{a}_k = \pi(\mathbf{s}_{k}) $.
    
3) The transition probability $\operatorname{Pr}(\mathbf{s}_{k+1}|\mathbf{s}_k, \mathbf{a}_k)$ is the probability of the next state $\mathbf{s}_{k+1}$ given the current state $\mathbf{s}_{k}$ and the  action $\mathbf{a}_k$. 
Since we adopt stationary scheduling policies, the state transition is independent of the time index. Thus, we drop the subscript $k$ here and use $\mathbf{s}$ and $\mathbf{s}^+$ to represent the current and the next states, respectively.
Due to the i.i.d. fading channel states, we have 
$
    \operatorname{Pr}(\mathbf{s}^+|\mathbf{s}, \mathbf{a}) = \operatorname{Pr}(\bm{\tau}^+|\bm{\tau},\mathbf{H}, \mathbf{a}) \operatorname{Pr}(\mathbf{H}^+),
$
where $\operatorname{Pr}(\mathbf{H}^+)$ can be obtained from \eqref{eq:q} and  $\operatorname{Pr}(\bm{\tau}^+|\bm{\tau},\mathbf{H}, \mathbf{a}) = \prod_{n=1}^N \operatorname{Pr}(\tau^+_{n}|\tau_{n}, \mathbf{H}, a_{n})$, which is derived based on \eqref{eq:q} and \eqref{eq:tau}
\begin{align}
    & \operatorname{Pr}(\tau^+_{n}|\tau_{n}, \mathbf{H}, a_{n})  = \left\{
    \begin{array}{l}
        1 - p_{h_{n, m}}, \ \ \, \text{if $\tau^+_{n} = 1, a_{n} = m$} \\
        p_{h_{n, m}}, \qquad \; \; \text{if $\tau^+_{n} = \tau_{n} + 1, a_{n} = m$}\\
        1, \qquad \quad \ \  \text{\ \ if $\tau^+_{n} = \tau_{n} + 1, a_{n} = 0$} \\
        0, \qquad \quad \ \ {\ \ \text{otherwise.}}
    \end{array}
    \right.
\end{align}

4) The immediate reward of Problem~\ref{pro1} at time $k$ is defined as $\sum_{n=1}^N -\text{Tr} (\mathbf{P}_{n,k})$. Since $\mathbf{P}_{n,k}$ defined in \eqref{eq:MSE} is a function of $\tau_{n,k}$, the reward is represented as $r(\mathbf{s}_k)$, \emph{monotonically decreasing with the AoI state}.

\subsection{Threshold Policies}
	Many works have proved that optimal scheduling policies have the threshold structure in Definition~\ref{def: threshold} for different problems, where the reward of an individual user is a monotonic function of its state~\cite{wu2018optimalmulti, wu2020optimalmulti, hsu2017threshold}. 
	Our problem has the same property, where the estimation MSE of sensor $n$ decreases with the decreasing AoI state and the increasing channel state.
	
\begin{figure}[t]
    \centering
    \begin{subfigure}[Schedule actions at AoI states]
        {
        \begin{minipage}[t]{0.5\linewidth}
        \centering
        \includegraphics[width=\textwidth]{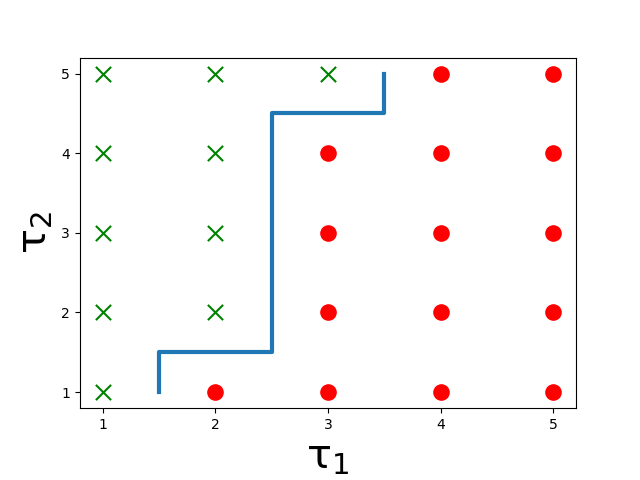}
        \end{minipage}
        \label{fig: threshold AoI}
        }
    \end{subfigure}
    \hspace{-0.9cm}
    \begin{subfigure}[Schedule actions at channel states]
        {
        \begin{minipage}[t]{0.5\linewidth}
        \centering
        \includegraphics[width=1\textwidth]{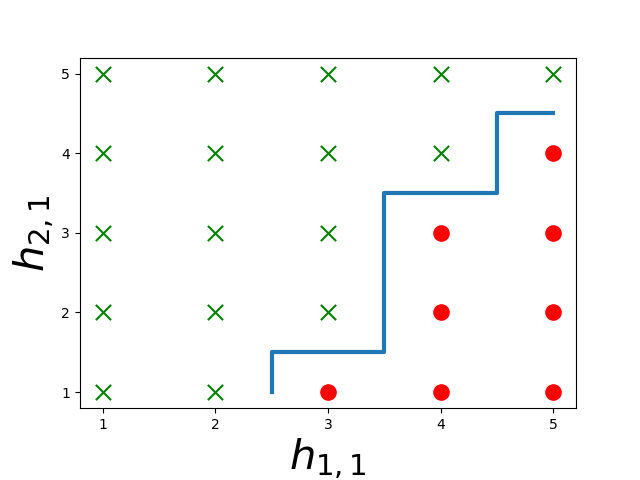}
        \end{minipage}
        \label{fig: threshold channel}
        }
    \end{subfigure}
    \vspace{-0.5cm}
    \caption{Structure of the optimal scheduling policy with $N=2$ and $M=1$, where $\bullet$ and $\times$ represent the schedule of sensor 1 and 2, respectively.}
    \vspace{-0.6cm}
    \label{fig: threshold}
\end{figure}

	To verify whether the optimal policy of Problem~\ref{pro1} has the threshold structure, we consider a system with $N=2$ and $M=1$ and adopt the conventional value iteration algorithm to find the optimal policy. As illustrated in Fig.~\ref{fig: threshold}, we see action switching curves in both AoI and channel state spaces, and the properties in Definition~\ref{def: threshold} are observed. Due to space limitations, we here do not present formal proof that the optimal policy has a threshold structure, but will investigate it in our future work. In Section~\ref{sec: simulation}, our numerical results also show that the optimized threshold policy is near optimal, confirming the conjecture.

\begin{definition}\label{def: threshold}
	For a threshold policy $\pi$ of Problem~\ref{pro1}, if channel~$m$ is assigned to sensor~$n$ at the state $\mathbf{s} = (\bm{\tau}, \mathbf{H})$, then the following two properties must hold:
	
	(i) for state $\mathbf{s}' = (\bm{\tau}, \mathbf{H}'_{n,m})$, where $\mathbf{H}'_{n,m}$ is identical to $\mathbf{H}$ except the sensor-$n$-channel-$m$ state with $h'_{n,m} > h_{n,m}$, then channel $m$ is still assigned to sensor~$n$;

	(ii) for state $\mathbf{s}' = (\bm{\tau}'_{(n)}, \mathbf{H})$, where $ \bm{\tau}'_{(n)}$ is idential to $\bm{\tau}$ except sensor $n$'s AoI with $\tau'_n>\tau_n$, then either channel $m$ or a better channel is assigned to sensor~$n$.
\end{definition}
For property (i), given that sensor $n$ is scheduled at channel $m$ at a certain state, if the channel quality of $h_{n,m}$ improves while the AoI and the other channel states are the same, then the threshold policy still assigns channel $m$ to sensor $n$. The property is reasonable since the channel quality of sensor $n$ to channel $m$ is the only factor changed and improved.

For property (ii), if the AoI state of sensor $n$ is increased while the other states remain the same, the  policy must schedule sensor $n$ to a channel that is no worse than the previous one. 
Such a scheduling policy does make sense, since a larger sensor $n$'s AoI leads to a lower reward, requiring a better channel for transmission to improve the reward effectively.

In the following, we will develop novel schemes that leverage the threshold structure to guide DRL algorithms for solving Problem~\ref{pro1} with improved training speed and performance.

\section{Structure-Enhanced DRL} \label{sec: DRL}
In the literature, DQN and DDPG are the most commonly adopted off-policy DRL algorithms for solving optimal scheduling problems (see \cite{leong2020DRL,pang2022drl} and references therein), and they provide significant performance improvements over heuristic policies.
Next, we develop threshold structure-enhanced (SE) DQN and DDPG for solving Problem~\ref{pro1}.
\subsection{Structure-Enhanced DQN}
We define the Q-value as the expected discounted sum of the reward function
\begin{equation}\label{eq: Q}
    Q^\pi(\mathbf{s}_k, \mathbf{a}_k) =  \mathbb{E} \left[\sum_{t=k}^{\infty} \gamma^{t-k} r(\mathbf{s}_t) \big \vert \mathbf{a}_k, \mathbf{a}_t = \pi(\mathbf{s}_t),\forall t>k \right],
\end{equation}
which measures the long-term performance of policy $\pi$ given the current state-action pair $(\mathbf{s}_k,\mathbf{a}_k)$. Let $\pi^*$ denote the optimal policy of Problem~\ref{pro1}. Then, the corresponding Q-value can, in principle, be obtained by solving the Bellman equation~\cite{mnih2013DQN}
\begin{equation}\label{eq:optimal Q}
    Q^{*}(\mathbf{s}_k, \mathbf{a}_k) = r(\mathbf{s}_k) + \mathbb{E}_{\mathbf{s}_{k+1}} \left[ \gamma \max_{\mathbf{a}_{k+1}}Q^{*} \left(\mathbf{s}_{k+1}, \mathbf{a}_{k+1}\right) \right].
\end{equation}
From the definition~\eqref{eq: Q}, the optimal policy generates the best action achieving the highest Q-value, and can be written as
\begin{equation}\label{eq:optimal Q action}
    \mathbf{a}^*_k = \pi^*(\mathbf{s}_k) = \mathop{\arg\max}_{\mathbf{a}_k}Q^{*}(\mathbf{s}_k, \mathbf{a}_k).
\end{equation}

A well-trained DQN uses a neural network (NN) with the parameter set $\bm{\theta}$ to approximate the Q-values of the optimal policy in~\eqref{eq:optimal Q action} as $Q(\mathbf{s}_k, \mathbf{a}_k;\bm{\theta})$ and use it to find the optimal action. 
Considering the action space defined in Section~\ref{sec:MDP}, the DQN has $N!/(N-M)!$ Q-value outputs of different state-action pairs.
To train the DQN, one needs to sample data (consisting of states, actions, rewards, and next states), define a loss function of $\bm \theta$ based on the collected data, and minimize the loss function to find the optimized $\bm \theta$. 
However, the conventional DQN training method has never utilized structures of optimal policies before.

To utilize the knowledge of the threshold policy structure for enhancing the DQN training performance, we propose \textbf{1) an SE action selection method} based on Definition~\ref{def: threshold} to select reasonable actions and hence  enhance the data sampling efficiency; 
and \textbf{2) an SE loss function definition} to add the penalty to sampled actions that do not follow the threshold structure.

Our SE-DQN training algorithm has three stages: 1) the DQN with \textbf{loose SE action selection} stage, which only utilizes part of the structural property,  2) the DQN with \textbf{tight SE action selection} stage utilizes the full structural property, and 3) the conventional DQN stage. 
The first two stages use the SE action selection schemes and the SE loss function to train the DQN fast, resulting in a reasonable threshold policy, and the last stage is for further policy exploration.
In what follows, we present the loose and tight SE action selection schemes and the SE loss function.

\subsubsection{Loose SE action selection}
We randomly select an action $\mathbf{a}^\epsilon$ from the entire action space with a probability of $\epsilon$ for action exploration; with a probability of $(1-\epsilon)$, we generate the SE action $\hat{\mathbf{a}}$ as below. For simplicity, we drop the time index when describing action selections.

The threshold structure suggests that the actions of $\mathbf{s}$ and the state with a smaller AoI or channel state are correlated. 
Thus, one can infer the action based on the action at the state with a smaller channel, or AoI state, based on property (i), or (ii), of Definition~\ref{def: threshold}, respectively.
We only consider property (ii) for loose SE action selection, as it is difficult to find actions that satisfy both structural properties (i) and (ii) at the beginning of training.  We will utilize both (i) and (ii) in the tight SE action selection stage.

Given the state $\mathbf{s} = (\bm{\tau}, \mathbf{H})$, we define the state with a smaller sensor $n$'s AoI as $\dot{\mathbf{s}}^{(n)} = (\dot{\bm{\tau}}^{(n)}, \mathbf{H})$, where 
\begin{equation}\label{eq: last AoI}
    \dot{\bm{\tau}}^{(n)} = (\tau_1, \dots, \tau_{n}-1, \dots, \tau_N).
\end{equation}
For each $n$, we calculate the corresponding action based on the Q-values as
\begin{align}\label{eq:AoI action}
    \dot{\mathbf{a}}^{(n)} \triangleq  (\dot{a}^{(n)}_{1}, \dots, \dot{a}^{(n)}_{n}, \dots, \dot{a}^{(n)}_{N})= \mathop{\arg\max}_{\dot{\mathbf{a}}} Q(\dot{\mathbf{s}}^{(n)}, \dot{\mathbf{a}}; \bm{\theta}).\
\end{align}
Recall that $\dot{a}^{(n)}_{n}\in\{0,1,\dots,M\}$ is the channel index assigned to sensor $n$ at the state $\dot{\mathbf{s}}^{(n)}$.

If $\dot{a}^{(n)}_{n}>0$, then property (ii) implies that channel $\dot{a}^{(n)}_{n}$ or a better channel is assigned to sensor $n$ at the state $\mathbf{s}$. We define the set of channels with better quality as
\begin{equation}\label{eq:set M}
\mathcal{M}^{(n)} = \{m'|h_{n,m'} > h_{n,\dot{a}^{(n)}_{n}}, m'=1,\dots,M \}.
\end{equation}
Then, the SE action for sensor $n$, say $\hat{a}_n$, is randomly chosen from the set $\mathcal{M}^{(n)}$ with probability $\xi$, and is equal to $\dot{a}^{(n)}_{n}$ with probability $1-\xi$.

If $\dot{a}^{(n)}_{n}=0$, then property (ii) can not help with determining the action. Then, we define the action selected by the greedy policy (i.e., the conventional DQN method) at the state $\mathbf{s}$ as
\begin{equation}\label{eq: greedy action}
\tilde{\mathbf{a}} \triangleq (\tilde{a}_1,\tilde{a}_2,\dots,\tilde{a}_N) = \mathop{\arg\max}_{\mathbf{a}} Q(\mathbf{s}, \mathbf{a}; \bm{\theta}).
\end{equation}
Thus, we set the SE action of sensor $n$ identical to the one generated by the conventional DQN method, i.e., $\hat{a}_n=\tilde{a}_n$.

Now we can define the SE action for $N$ sensors as $\hat{\mathbf{a}} = (\hat{a}_{1}, \hat{a}_{2}, \dots, \hat{a}_{N})$. If such an action meets the constraint
\begin{align}\label{eq: loose action constraint}
    \sum_{m=1}^{M} \boldsymbol{\mathbbm{1}}\left( \hat{a}_{n} = m \right) \leq 1, \quad
    \sum_{n=1}^{N} \boldsymbol{\mathbbm{1}}\left( \hat{a}_{n} = m \right) \leq 1,
\end{align}
which is less restrictive than \eqref{eq: action constraint},
we select the action $\mathbf{a}$ as $\hat{\mathbf{a}}$ and assign the unused channels randomly to unscheduled sensors; otherwise, $\mathbf{a}$ is identical to that of the conventional method as~$\tilde{\mathbf{a}}$. 

\subsubsection{Tight SE action selection} By using the loose SE action selection, we first infer the scheduling action of sensor $n$ at the state $\mathbf{s}$ based on the action of the state with a smaller AoI, $\dot{\mathbf{s}}^{(n)}$. Then, we check whether the loose SE action satisfies the threshold property (i) as below.

For notation simplicity, we use $m>0$ to denote the SE channel selection for sensor $n$. 
Given the state $\mathbf{s}$, we define the state with a smaller channel $m$'s state for sensor $n$ as $\ddot{\mathbf{s}}^{(n,m)} = (\bm{\tau}, \ddot{\mathbf{H}}^{(n,m)})$, where $\ddot{\mathbf{H}}^{(n,m)}$ and $\mathbf{H}$ are identical except the element $\ddot{h}^{(n,m)}_{n,m} = h_{n,m} -1$. Then, we calculate the corresponding action
\begin{equation} \label{eq:channel action}
    \ddot{\mathbf{a}}^{(n,m)} \!\!\triangleq\! (\ddot{a}^{(n,m)}_{1}\!\!\!, \!\dots,\! \ddot{a}^{(n,m)}_{n}\!\!\!, \dots, \ddot{a}^{(n,m)}_{N}) 
    \!=\! \mathop{\arg\max}_{\ddot{\mathbf{a}}} Q(\ddot{\mathbf{s}}^{(n,m)}\!\!\!, \ddot{\mathbf{a}};\! \bm{\theta}). 
\end{equation}
From the structural properties (i) and (ii), the scheduling action $\ddot{a}^{(n,m)}_{n}$ should be identical to $m$. 
Thus, if $\ddot{a}^{(n,m)}_{n}=m$, then the SE action for sensor $n$ is $\hat{a}_n=m$; otherwise, $\hat{a}_n$ is identical to the conventional DQN action.
The SE action satisfying the structural properties (i) and (ii) is $\hat{\mathbf{a}} = (\hat{a}_{1}, \hat{a}_{2}, \dots, \hat{a}_{N})$. If such an action meets the constraint~\eqref{eq: action constraint}, 
then it is executed as $\mathbf{a} = \hat{\mathbf{a}}$; otherwise, we select the greedy action $\mathbf{a} = \tilde{\mathbf{a}}$.

\subsubsection{SE loss function}
During the training, each transition $(\mathbf{s}_t, \mathbf{a}_t, \hat{\mathbf{a}}_t, \tilde{\mathbf{a}}_{t}, r_t, \mathbf{s}_{t+1})$ is stored in a replay memory, where $r_t\triangleq r(\mathbf{s}_t)$ denotes the immediate reward.
Different from the conventional DQN, we include both the SE action $\hat{\mathbf{a}}$ and the greedy action $\tilde{\mathbf{a}}$, in addition to the executed action $\mathbf{a}$.

Let $\mathcal{T}_{i} \triangleq (\mathbf{s}_i, \mathbf{a}_i, \hat{\mathbf{a}}_i, \tilde{\mathbf{a}}_{t}, r_i, \mathbf{s}_{i+1})$ denote the $i$th transition of a sampled batch from the replay memory. 
Same as the conventional DQN, we define the temporal-difference (TD) error of $\mathcal{T}_i$ as
\begin{equation}\label{eq:TD}
    \mathsf{TD}_{i} \triangleq y_{i}-Q(\mathbf{s}_{i},\mathbf{a}_{i};\bm{\theta}),
\end{equation} 
where $y_{i} = r_{i} + \gamma \max_{{\mathbf{a}}_{i+1}} Q(\mathbf{s}_{i+1},{\mathbf{a}}_{i+1};\bm{\theta})$ is the estimation of Q-value at next step.
This is to measure the gap between the left and right sides of the Bellman equation~\eqref{eq:optimal Q}. A larger gap indicates that the approximated Q-values are far from the optimal.
Different from DQN, we introduce the action-difference (AD) error as below to measure the difference of Q-value between actions selected by the SE strategy and the greedy strategy:
\begin{equation}\label{eq:AD}
    \mathsf{AD}_{i} \triangleq Q(\mathbf{s}_{i},\hat{\mathbf{a}}_{i};\bm{\theta}) - Q(\mathbf{s}_{i}, \tilde{\mathbf{a}}_{i};\bm{\theta}).
\end{equation}
Since the optimal policy has the threshold structure, the inferred action $\hat{\mathbf{a}}$ should be identical to the optimal action $\tilde{\mathbf{a}}$. Thus, a well-trained $\bm \theta$ should lead to a small difference in \eqref{eq:AD}.

Based on \eqref{eq:TD} and \eqref{eq:AD}, we define the loss function of $\mathcal{T}_i$ as
\begin{equation}\label{loss:critic_i}
    L_{i}(\bm{\theta}) = \left\{
    \begin{array}{l}
    \begin{aligned}
        & \alpha_1 \mathsf{TD}^{2}_{i} + (1-\alpha_1) \mathsf{AD}^{2}_{i},  \quad \text{if $\mathbf{a}_{i} = \hat{\mathbf{a}}_{i}$} \\
        & \mathsf{TD}^{2}_{i}, \qquad \qquad \qquad \qquad \ \; \,  \text{otherwise},
    \end{aligned}
    \end{array}
    \right. 
\end{equation}
where $\alpha_1$ is a hyperparameter to balance the importance of $\mathsf{TD}_i$ and $\mathsf{AD}_i$.
In other words, if the SE action is executed, both the TD and AD errors are taken into account; otherwise, the conventional TD-error-based loss function is adopted.

Given the batch size $B$, the overall loss function is
\begin{equation}\label{loss:critic}
    L(\bm{\theta}) = \frac{1}{B} \sum_{i=1}^{B} L_{i}(\bm{\theta}).
\end{equation}
To optimize $\bm \theta$, we adopt the well-known gradient descent method and calculate the gradient as below
\begin{equation}\label{grat:critic}
    \nabla_{\bm{\theta}} L(\bm{\theta}) = \frac{1}{B} \sum_{i=1}^{B} \nabla_{\bm{\theta}} L_{i}(\bm{\theta})
\end{equation}
where $\nabla_{\bm{\theta}} L_{i}(\bm{\theta})$ is given as
\begin{equation}\label{grat:critic_i}
    \nabla_{\bm{\theta}} L_{i}(\bm{\theta}) = \left\{
    \begin{array}{l}
    \begin{aligned}
        & \!\!\!-2 ((1-\alpha_1) \mathsf{AD}_{i} \nabla_{\bm{\theta}}\left(Q\left(\mathbf{s}_{i},\hat{\mathbf{a}}_{i};\bm{\theta}) - Q(\mathbf{s}_{i}, \tilde{\mathbf{a}}_{i};\bm{\theta}\right)\right) \\
        & \ \ +\alpha_1\mathsf{TD}_{i} \nabla_{\bm{\theta}} Q(\mathbf{s}_{i}, \mathbf{a}_{i};\bm{\theta})), \quad \text{if $\mathbf{a}_{i} = \hat{\mathbf{a}}_{i}$} \\
        & \!\!\! -2(\mathsf{TD}_{i} \nabla_{\bm{\theta}} Q(\mathbf{s}_{i}, \mathbf{a}_{i};\bm{\theta})), \qquad \; \, \text{otherwise}.
    \end{aligned}
    \end{array}
    \right. 
\end{equation}

The details of the SE-DQN algorithm are given in Algorithm~\ref{alg:SE-DQN}. 


\begin{algorithm}[t]
    \small
    \caption{\small{SE-DQN for sensor scheduling in the  remote estimation system}}
    \label{alg:SE-DQN}
    \begin{algorithmic}[1]
        \State Initialize replay memory $\mathcal{D}$ to capacity $N$
        \State Initialize policy network with random weights $\bm{\theta}$
        \State Initialize target network with weights $\hat{\bm{\theta}} = \bm{\theta}$
        \For {episode $= 1, 2, \dots, E_{1}$}
            \State Initialize state $\mathbf{s}_0$
            \For {$t = 1, 2, \dots, T$}
                \State Generate $\hat{\mathbf{a}}_t$ and select action $\mathbf{a}_t$ by the loose SE action selection method of SE-DQN
                \State Execute action $\mathbf{a}_t$ and observe $r_t$ and $\mathbf{s}_{t+1}$
                \State Compute action $\tilde{\mathbf{a}}_{t} = \mathop{\arg\max}_{\mathbf{a}} Q(\mathbf{s}_{t}, \mathbf{a}; \bm{\theta})$
                \State Store transition $(\mathbf{s}_t, \mathbf{a}_t, \hat{\mathbf{a}}_t, \tilde{\mathbf{a}}_{t}, r_t, \mathbf{s}_{t+1})$ in $\mathcal{D}$
                \State Sample a random batch of transitions $(\mathbf{s}_{i}, \mathbf{a}_{i}, \hat{\mathbf{a}}_{i}, \tilde{\mathbf{a}}_{i}, r_{i}, \mathbf{s}_{i+1})$
                \State Calculate $\mathsf{TD}_i$ and $\mathsf{AD}_i$ based on \eqref{eq:TD} and \eqref{eq:AD}
                \State Update $\bm \theta$ according to equation \eqref{grat:critic}
                \State Every $t'$ steps set $\hat{\bm{\theta}} = \bm{\theta}_{t}$
            \EndFor
        \EndFor
        \For {episode $= E_{1}, \dots, E_{2}$}
            \State Repeat algorithm from line 5 to line 15 by adopting the tight SE action selection method
        \EndFor
        \For {episode $= E_{2}, \dots, E$}
            \State Execute the DQN algorithm as in \cite{leong2020DRL}
        \EndFor
    \end{algorithmic}
\end{algorithm}

\subsection{Structure-Enhanced DDPG}
Different from the DQN, which has one NN to estimate the Q-value, a DDPG agent has two NNs~\cite{lillicrap2015DDPG}, a critic NN with parameter $\bm \theta$ and an actor NN with the parameter $\bm \mu$.
In particular,  the actor NN   approximates the optimal policy $\pi^*(\mathbf{s})$ by $\mu(\mathbf{s};\bm{\mu})$, while the critic NN approximates the Q-value of the optimal policy $Q^*(\mathbf{s}, \mathbf{a})$ by $Q(\mathbf{s}, \mathbf{a};\bm{\theta})$.
In general, the critic NN judges whether the generated action of the actor NN is good or not, and the latter can be improved based on the former. The critic NN is updated by minimizing the TD error similar to the DQN.
The actor-critic framework enables DDPG to solve MDPs with continuous and large action space, which cannot be handled by the DQN.

To solve our scheduling problem with a discrete action space, we adopt an action mapping scheme similar to the one adopted in~\cite{pang2022drl}. 
We set the direct output of the actor NN with $N$ continuous values, ranging from $-1$ to $1$, corresponding to sensors $1$ to $N$, respectively. Recall that the DQN has $N!/(N-M)!$ outputs.
The $N$ values are sorted in descending order. The sensors with the highest $M$ ranking are assigned to channels $1$ to $M$, respectively. The corresponding ranking values are then linearly normalized to $[-1,1]$ as the output of the final outputs of the actor NN, named as the virtual action $\mathbf{v}$. 
It directly follows that the virtual action $\mathbf{v}$ and the real scheduling action $\mathbf{a}$ can be mapped from one to the other directly. 
Therefore, we use the virtual action $\mathbf{v}$, instead of the real action $\mathbf{a}$, when presenting the SE-DDPG algorithm.

Similar to the SE-DQN, the SE-DDPG has the loose SE-DDPG stage, the tight SE-DDPG stage, and the conventional DDPG stage. The $i$th sampled transition is denoted as
\begin{equation}
\mathcal{T}_{i} \triangleq (\mathbf{s}_i, \mathbf{v}_i, \hat{\mathbf{v}}_i, \tilde{\mathbf{v}}_{t}, r_i, \mathbf{s}_{i+1}).
\end{equation} 
We present the SE action selection method and the SE loss function in the sequel.

\subsubsection{SE action selection}\label{subsec:SE-DDPG action} The general action selection approach for the SE-DDPG is identical to that of the SE-DQN, by simply converting $\mathbf{v}_i$, $\mathbf{v}^\epsilon_i$, $\tilde{\mathbf{v}}_i$, and $\hat{\mathbf{v}}_i$ to $\mathbf{a}_i$, $\mathbf{a}^\epsilon_i$, $\tilde{\mathbf{a}}_i$, and $\hat{\mathbf{a}}_i$, respectively. 
Different from DQN with $\epsilon$-greedy actions, the action generated by the DDPG based on the current state is 
\begin{equation}\label{eq:tilde_v}
\tilde{\mathbf{v}}_i = \mu(\mathbf{s}_i;\bm{\mu}),
\end{equation}
and the random action $\mathbf{v}^\epsilon_i$ was generated by adding noise to the original continuous output of the actor NN.

\begin{algorithm}[t]
\small
    \caption{\small{SE-DDPG for sensor scheduling in the  remote estimation system}}
    \label{alg:SE-DDPG}
    \begin{algorithmic}[1]
        \State Initialize replay memory $\mathcal{D}$ to capacity $N$
        \State Initialize actor network and critic network with random weights $\bm{\mu}$ and $\bm{\theta}$
        \State Initialize target network and with weight $\hat{\bm{\mu}} = \bm{\mu}$, $\hat{\bm{\theta}} = \bm{\theta}$
        \For {episode $= 1, 2, \dots, E_{1}$}
            \State Initialize a random noise $\mathcal{N}$ for action exploration
            \State Initialize state $\mathbf{s}_0$
            \For {$t=1, 2, \dots, T$}
                \State Generate $\hat{\mathbf{v}}_{t}$ and select action $\mathbf{v}_{t}$ by the loose SE action selection method of SE-DDPG
                \State Mapping virtual action $\mathbf{v}_{t}$ to real action $\mathbf{a}_{t}$
                \State Execute action $\mathbf{a}_{t}$ and observe $r_t$ and $\mathbf{s}_{t+1}$
                \State Compute action $\tilde{\mathbf{v}}_{t} = \mu(\mathbf{s}_t;\bm{\mu})$
                \State Store transition $(\mathbf{s}_t, \mathbf{v}_t, \hat{\mathbf{v}}_t, \tilde{\mathbf{v}}_{t}, r_t, \mathbf{s}_{t+1})$ in $\mathcal{D}$
                \State Sample a random batch of transitions $(\mathbf{s}_{i}, \mathbf{v}_{i}, \hat{\mathbf{v}}_{i}, \tilde{\mathbf{v}}_{i}, r_{i}, \mathbf{s}_{i+1})$
                \State Calculate $\mathsf{TD}_i$ and $\mathsf{AD}_i$ based on \eqref{eq:TD} and \eqref{eq:AD} but with the virtual actions $\mathbf{v}_{i}, \hat{\mathbf{v}}_{i}, \Tilde{\mathbf{v}}_{i}$
                \State Update $\bm{\theta}$ according to equation \eqref{grat:critic}
                \State Update $\bm{\mu}$ according to equation \eqref{grat:actor}
                \State Update the target network:
                \begin{align}
                    \hat{\bm{\mu}} & \gets \delta \bm{\mu} + (1 - \delta)\hat{\bm{\mu}} \\
                    \hat{\bm{\theta}} & \gets \delta \bm{\theta} + (1 - \delta)\hat{\bm{\theta}}
                \end{align}
            \EndFor
        \EndFor
        \For {episode $E_{1} = 1, 2, \dots, E_{2}$}
            \State Repeat algorithm from line 5 to line 18 by adopting the tight SE action selection method
        \EndFor
        \For {episode $E_{2} = 1, 2, \dots, E$}
            \State Execute the conventional DDPG algorithm as in \cite{lillicrap2015DDPG}
        \EndFor
    \end{algorithmic}
\end{algorithm}

\subsubsection{SE loss function} Different from the SE-DQN, the SE-DDPG needs different loss functions for updating the critic NN and the actor NN.
For the critic NN, we use the same loss function as in \eqref{loss:critic}, and thus the gradient for the critic NN update is identical to \eqref{grat:critic}.
Note that for DDPG, the next virtual action $\tilde{\mathbf{v}}_{i+1}$ is the direct output of the actor NN given the next state $\mathbf{s}_{i+1}$, i.e., $\tilde{\mathbf{v}}_{i+1} = \mu(\mathbf{s}_{i+1};\bm \mu)$. Thus, when calculating the TD error \eqref{eq:TD}, we have $y_{i} = r_{i} + \gamma Q(\mathbf{s}_{i+1}, {\mu}(\mathbf{s}_{i+1};\bm\mu);\bm{\theta})$.

For the actor NN, we introduce the difference between actions selected by the SE strategy $\hat{\mathbf{v}}_i$ and the actor NN  $\tilde{\mathbf{v}}_i$, when $\hat{\mathbf{v}}_i$ is executed, i.e., $\mathbf{v}_i=\hat{\mathbf{v}}_i$. If the SE action is not selected, then the loss function is the Q-value given the state-action pair, which is identical to the conventional DDPG. Given the hyperparameter $\alpha_2$, the loss function for the transition~$\mathcal{T}_i$ is defined as
\begin{equation}\label{loss:actor_i}
    L_{i}(\bm{\mu}) = \left\{
    \begin{array}{l}
    \begin{aligned}
        & \alpha_2 Q(\mathbf{s}_{i}, \tilde{\mathbf{v}}_{i}; \bm{\theta}) + (1-\alpha_2) \left( \mathbf{v}_{i} - \tilde{\mathbf{v}}_{i} \right)^{2},  \ \ \text{if $\mathbf{v}_{i} = \hat{\mathbf{v}}_{i}$} \\
        & Q(\mathbf{s}_{i}, \tilde{\mathbf{v}}_{i}; \bm{\theta}), \qquad \qquad \qquad \qquad \qquad \ \ \ \, \text{otherwise},
    \end{aligned}
    \end{array}
    \right. 
\end{equation}
and hence the overall loss function given the sampled batch is 
\begin{equation}\label{loss:actor}
    L(\bm{\mu}) = \frac{1}{B} \sum_{i=1}^{B} L_{i}(\bm{\mu}).
\end{equation}
By replacing \eqref{eq:tilde_v} and \eqref{loss:actor_i} in \eqref{loss:actor} and applying the chain rule, we can derive the gradient of the overall loss function in terms of $\bm \mu$ as
\begin{equation}\label{grat:actor}
    \nabla_{\bm{\mu}} L(\bm{\mu}) = \frac{1}{B} \sum_{i=1}^{B} \nabla_{\bm{\mu}} L_{i}(\bm{\mu}),
\end{equation}
where $\nabla_{\bm{\mu}} L_{i}(\bm{\mu})$ is given by:
\begin{equation}\label{grat:actor_i}
    \nabla_{\bm{\mu}} L_{i}(\bm{\mu}) = \left\{
    \begin{array}{l}
    \begin{aligned}
        & \alpha_2 \nabla_{\tilde{\mathbf{v}}_{i}} Q(\mathbf{s}_{i}, \tilde{\mathbf{v}}_{i}; \bm{\theta}) \nabla_{\bm{\mu}} \mu(\mathbf{s}_{i};\bm{\mu}) - 2 (1-\alpha_2) \\
        & \ \ \left( \mathbf{v}_{i} - \mu\left(\mathbf{s}_{i};\bm{\mu}\right) \right) \nabla_{\bm{\mu}} \mu(\mathbf{s}_{i};\bm{\mu}) , \quad \text{if $\mathbf{v}_{i} = \hat{\mathbf{v}}_{i}$} \\
        & \nabla_{\tilde{\mathbf{v}}_{i}} Q(\mathbf{s}_{i}, \tilde{\mathbf{v}}_{i}; \bm{\theta}) \nabla_{\bm{\mu}} \mu(\mathbf{s}_{i};\bm{\mu}), \qquad \text{otherwise}.
    \end{aligned}
    \end{array}
    \right. 
\end{equation}

The details of the proposed  SE-DDPG algorithm are given in Algorithm~\ref{alg:SE-DDPG}.

\section{Numerical Experiments} \label{sec: simulation}
\begin{table}[t]
    \footnotesize
    \setlength\tabcolsep{7.5pt}
    \centering
    \caption{Summary of Hyperparameters}
    \setlength{\tabcolsep}{0.5mm}
    \label{tab:Setup}
    \vspace{-0.3cm}
    \begin{tabular}{c|c}
         \hline \hline
         \textbf{Hyperparameters of SE-DQN and SE-DDPG} & Value \\
         \hline
         Initial values of $\epsilon$ and $\xi$ & 1 \\
         \rowcolor[HTML]{EFEFEF} 
         Decay rates of $\epsilon$ and $\xi$                     & 0.999 \\
         Minimum $\epsilon$ and $\xi$                        & 0.01 \\
         \rowcolor[HTML]{EFEFEF} 
         Mini-batch size, $B$                       & 128 \\
         Experience replay memory size, $K$         & 20000 \\
         \rowcolor[HTML]{EFEFEF} 
         Discount factor, $\gamma$                  & 0.95 \\
         Weight of SE-DQN and critic network loss function, $\alpha_1$ & 0.5 \\
         \hline
         \textbf{Hyperparameters of SE-DQN} \\
         \hline
         Learning rate                              & 0.0001 \\
         \rowcolor[HTML]{EFEFEF}
         Decay rate of learning rate                & 0.001 \\
         Target network update frequency            & 100 \\
         \rowcolor[HTML]{EFEFEF}
         Input dimension of network, $Q(\mathbf{s}, \mathbf{a};\bm{\theta})$   & $N+N \times M$ \\
         Output dimension of network, $Q(\mathbf{s}, \mathbf{a};\bm{\theta})$      & $N!/(N-M)!$ \\
         \hline
         \textbf{Hyperparameters of SE-DDPG} \\
         \hline
         Learning rate of actor network             & 0.0001 \\
         \rowcolor[HTML]{EFEFEF}
         Learning rate of critic network             & 0.001 \\
         Decay rate of learning rate                & 0.001 \\
         \rowcolor[HTML]{EFEFEF}
         Soft parameter for target update, $\delta$ & 0.005 \\
         Weight of actor network loss function, $\alpha_2$ & 0.9 \\
         \rowcolor[HTML]{EFEFEF}
         Input dimension of actor network, $\mu(\mathbf{s}_t;\bm{\mu})$   & $N+N \times M$ \\
         Output dimension of actor network, $\mu(\mathbf{s}_t;\bm{\mu})$   & $N$ \\
         \rowcolor[HTML]{EFEFEF}
         Input dimension of critic network, $Q(\mathbf{s}, \mathbf{a};\bm{\theta})$   & $2N+N \times M$ \\
         Output dimension of critic network, $Q(\mathbf{s}, \mathbf{a};\bm{\theta})$   & $1$ \\
         \hline \hline
    \end{tabular}
    \vspace{-0.5cm}
\end{table}

In this section, we evaluate and compare the performance of the proposed SE-DQN and SE-DDPG with the benchmark DQN and DDPG.
\subsection{Experiment Setups}
Our numerical experiments run on a computing
platform with an Intel Core i5 9400F CPU @ 2.9 GHz with 16GB RAM and an NVIDIA RTX 2070 GPU. For the remote estimation system, we set the dimensions of the process state and the measurement as $l_{n} = 2$ and $e_{n} = 1$, respectively. The system matrices $\{\mathbf{A}_{n}\}$ are randomly generated with the spectral radius within the range of $(1, 1.4)$. The entries of $\mathbf{C}_{n}$ are drawn uniformly from the range $(0, 1)$, $\mathbf{W}_n$ and $\mathbf{V}_n$ are identity matrices. The fading channel state is quantized into $\bar{h}=5$ levels, and the corresponding packet drop probabilities are $0.2, 0.15, 0.1, 0.05,$ and $0.01$. The distributions of the channel states of each sensor-channel pair $(n,m)$, i.e., $q^{(n,m)}_1,\dots,q^{(n,m)}_{\bar{h}}$ are generated randomly.

During the training, we use the ADAM optimizer for calculating the gradient and reset the environment for each episode with $T=500$ steps. The episode numbers for the loose SE action, the tight SE action, and the conventional DQN stages are 50, 100, and 150, respectively. The settings of the hyperparameters for Algorithm~\ref{alg:SE-DQN} and Algorithm~\ref{alg:SE-DDPG} are summarized in Table~\ref{tab:Setup}.


\subsection{Performance Comparison}
\begin{figure}[t]
    \centering
    \includegraphics[width=0.95\linewidth]{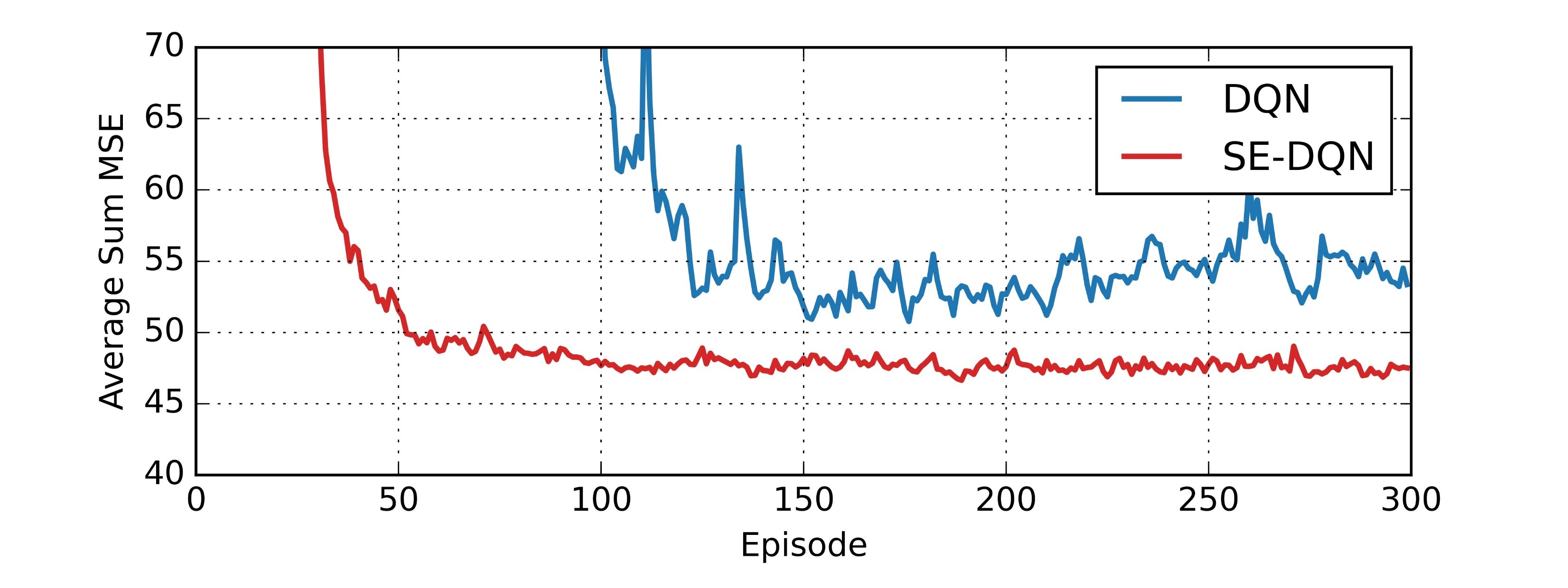}
    \vspace{-0.5cm}
    \caption{Average sum MSE of all processes during training with $N=6, M=3$.}
    \vspace{-0.4cm}
    \label{fig:training curve DQN}
\end{figure}

\begin{figure}[t]
    \centering
    \includegraphics[width=0.95\linewidth]{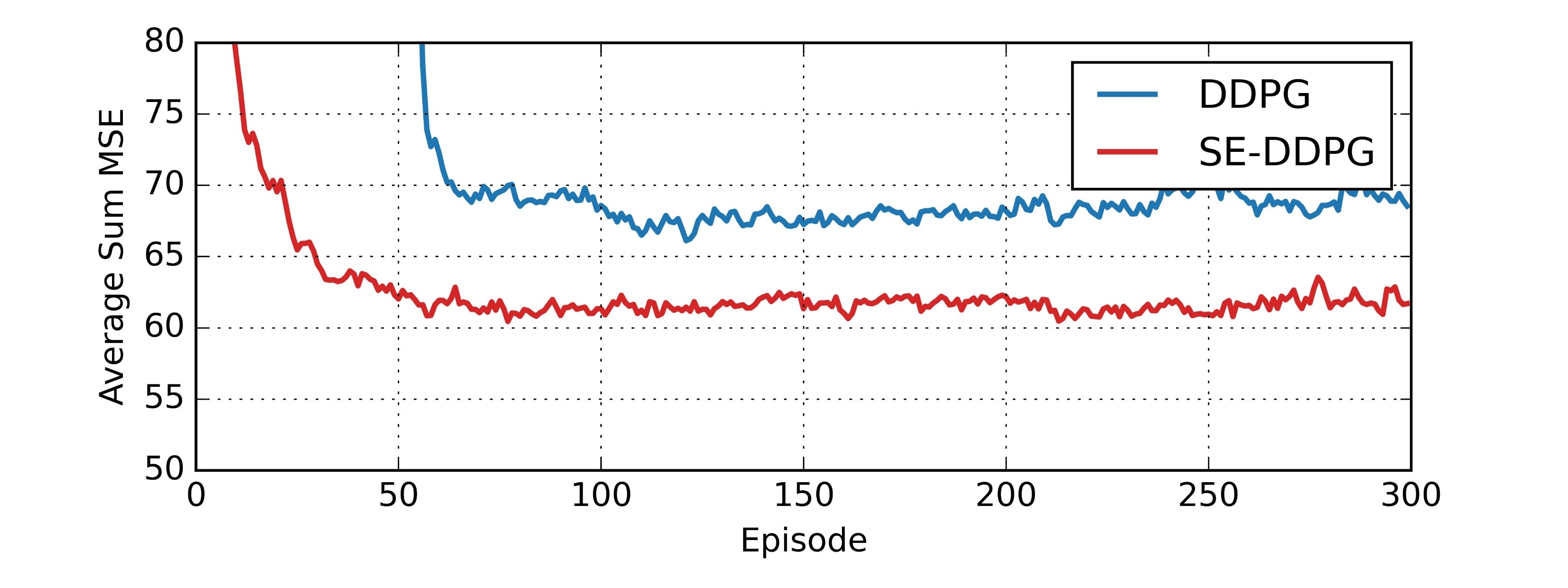}
    \vspace{-0.5cm}
    \caption{Average sum MSE of all processes during training with $N\!=10, M\!=5$.}
    \vspace{-0.4cm}
    \label{fig:training curve DDPG}
\end{figure}

Fig.~\ref{fig:training curve DQN} and Fig.~\ref{fig:training curve DDPG} illustrate the average sum MSE of all processes during the training achieved by the SE-DRL algorithms and the benchmarks under some system settings. Fig.~\ref{fig:training curve DQN} shows that the SE-DQN saves about $50\%$ training episodes for the convergence, and also decreases the average sum MSE for $10\%$ than the conventional DQN. 
Fig.~\ref{fig:training curve DDPG} shows that the SE-DDPG saves about $30\%$ training episodes and reduces the average sum MSE by $10\%$, when compared to the conventional DDPG. Also, we see that the conventional DQN and DDPG stages (i.e. the last 150 episodes) in Fig~\ref{fig:training curve DQN} and~\ref{fig:training curve DDPG} cannot improve much of the training performance. This implies that the SE stages have found near optimal policies.

In Table~\ref{tab:test result}, we test the performance of well-trained SE-DRL algorithms for different numbers of sensors and channels, and different settings of system parameters, i.e.,  $\mathbf{A}_{n}$, $\mathbf{C}_{n}$, and $q^{(n,m)}_1,\dots,q^{(n,m)}_{\bar{h}}$, based on 10000-step simulations. 
We see that for 6-sensor-3-channel systems, the SE-DQN reduces the average MSE by $10\%$ to $25\%$ over DQN, while both DDPG and SE-DDPG achieve similar performance as the SE-DQN. This suggests that the SE-DQN is almost optimal and that the DDPG and the SE-DDPG cannot improve further.
In 10-sensor-5-channel systems, neither the DQN nor the SE-DQN can converge, and the SE-DDPG achieves a $10\%$ MSE reduction over the DDPG. In particular, the SE-DDPG is the only converged algorithm in Experiment 12.
In 20-sensor-10-channel systems, we see that the SE-DDPG can reduce the average MSE by $15\%$ to $25\%$. Therefore, the performance improvement of the SE-DDPG appears significant for large systems.

\begin{table}[t]
	\footnotesize
	\setlength\tabcolsep{7.5pt}
	\centering
	\caption{Performance Comparison of the SE-DRL and the Benchmarks in terms of the Average Estimation MSE}
	\label{tab:test result}
	\setlength{\tabcolsep}{1.5mm}
	\vspace{-0.3cm}
	\begin{tabular}{c|c|c|c|c|c}
		\hline \hline
		\thead{System Scale \\ $(N,M)$} &  \thead{Experiment} & DQN & \textbf{SE-DQN} & DDPG & \textbf{SE-DDPG}\\
		\hline
		\rowcolor[HTML]{EFEFEF}
		$(6, 3)$ & 1 & 52.4121 & \textbf{47.6766} & 48.4075 & \textbf{47.0594} \\
		& 2 & 67.4247 & \textbf{49.8476} & 53.3675 & \textbf{44.7423} \\
		\rowcolor[HTML]{EFEFEF}
		& 3 & 84.1721 & \textbf{59.9127} & 56.9504 & \textbf{55.2409} \\
		& 4 & 79.5902 & \textbf{65.1640} & 64.4313 & \textbf{58.5534} \\
		\hline
		
		\rowcolor[HTML]{EFEFEF}
		
		\rowcolor[HTML]{EFEFEF}
		$(10, 5)$ & 9 & $-$ & $-$ & 68.0247 &  \textbf{62.4727} \\
		& 10 & $-$ & $-$ & 89.6290 & \textbf{78.4138} \\
		\rowcolor[HTML]{EFEFEF}
		& 11 & $-$ & $-$ & 90.2812 & \textbf{81.1148} \\
		& 12 & $-$ & $-$ & $-$ & \textbf{147.2844} \\
		\hline
		
		\rowcolor[HTML]{EFEFEF}
		$(20, 10)$ & 13 & $-$ & $-$ & 181.4135 & \textbf{159.4321} \\
		& 14 & $-$ & $-$ & 163.1257 & \textbf{142.1850} \\
		\rowcolor[HTML]{EFEFEF}
		& 15 & $-$ & $-$ & 173.0940 & \textbf{144.8166} \\
		& 16 & $-$ & $-$ & 215.2231 & \textbf{160.2304} \\
		\hline \hline
	\end{tabular}
 \vspace{-0.5cm}
\end{table}




\section{Conclusion}
We have developed the SE-DRL algorithms for optimal sensor scheduling of remote estimation systems. In particular, we have proposed novel SE action selection and loss functions to enhance training efficiency. 
Our numerical results have shown that the proposed structure-enhanced DRL algorithms can save the training time by 50\% and reduce the estimation MSE by 10\% to 25\%, when benchmarked against conventional DRL algorithms.
For future work, we will rigorously prove the threshold structure of the optimal scheduling policy for our system. Also, we will investigate other structural properties of the optimal policy to enhance the DRL.

\bibliographystyle{IEEEtran}

\end{document}